\documentclass{iopjournal}
\usepackage{fancyhdr}
\usepackage{amsmath}
\usepackage{amssymb}
\usepackage{graphicx}
\usepackage{dcolumn}
\usepackage{bm}
\usepackage{xcolor}
\usepackage{hyperref}
\usepackage{orcidlink}
\usepackage{cite}
\usepackage{bbold}
\usepackage{dsfont}

\newcommand{\integ}[3]{\int_{#1}^{#2}\text{d}#3~}

\newcommand{\Laplace}[1]{\mathcal{L}\{#1\}(s)}
\newcommand{\new}[1]{#1}

\begin{document}

\articletype{Paper} 

\title{Graph theoretic derivation of mutual 
linearity for transient probabilities and hitting
time distributions in Markov networks}

\author{Julian B. Voits \orcidlink{0009-0006-2650-6968} and Ulrich S. Schwarz\textsuperscript{*}\orcidlink{0000-0003-1483-640X}}

\affil{Institute for Theoretical Physics, Heidelberg University, Germany}

\affil{BioQuant-Center for Quantitative Biology, Heidelberg University, Germany}

\affil{$^*$Author to whom any correspondence should be addressed.}

\email{schwarz@thphys.uni-heidelberg.de}


\begin{abstract}
For irreducible, time-homogeneous Markov networks,
mutual linearity has recently been established for both occupation probabilities and network currents in the stationary regime as well as in the non-stationary regime in Laplace space. The derivation of this property for the stationary distribution utilized the Markov chain tree theorem, which also allows for an explicit combinatorial expression of the response ratios under variation of a single transition rate. The extension of this result was proven at the trajectory level by employing the Doob-Meyer decomposition. By employing the all-minors matrix-tree theorem, we show that this property also follows from a graph theoretic formulation and derive explicit combinatorial expressions for the non-stationary response ratios. The stationary result follows as the long-time limit. We also show that the small-time asymptotics are entirely determined by minimal path distances in the underlying graph. Finally we use the graph theoretic approach to prove that mutual linearity also extends to hitting time densities. 
\end{abstract}

\section{Introduction}
\label{sec:level1}

Markov networks describe stochastic dynamics of a process $(X_t)_{t\geq 0}$ on a discrete state space $\{1,...,N\}$ with memoryless transitions between states~\cite{gardiner2004handbook, norris1998markov}. Such processes are of central interest in probability theory and statistical physics~\cite{lyons2017probability,grimmett2020probability,van1992stochastic, Honerkamp}, but are also frequently applied in the modeling of complex biological and economic systems~\cite{gillespie1992rigorous,paul1999stochastic,bressloff2014stochastic}. A particularly interesting question is
when the stochastic process hits a certain threshold
for the first time, the so-called first passage
or first hitting time \cite{redner2001guide,metzler2014first}.

For a process in continuous time $t\geq0$ on a network with $N\in\mathbb{N}$ states, the occupation probabilities $p_n(t)=\mathbb{P}(X_t=n)$ of state $n\in\{1,...,N\}$ obey a linear, first-order differential equation, known as master equation or Kolmogorov forward equation~\cite{van1992stochastic}, which for time-homogeneous transition rates $k_{ij}$ from state $i$ to state $j$ reads:
\begin{equation}
    \dot{p}(t)=Wp(t),
    \label{eq:master_equation}
\end{equation}
where the infinitesimal generator $W$ is defined as:
\begin{align}
    W_{ij}:=\begin{cases}
        \qquad k_{ji}&\text{if}~i\neq j,\\-\sum_{l=1}^Nk_{il} &\text{if}~i= j.
    \end{cases}
\end{align}
\new{These dynamics can be related to graph theory because} 
$-W$ coincides with the definition of the out-degree Laplacian matrix of the edge-weighted directed graph $\mathcal{G}$ on $N$ vertices that is obtained from choosing the states as vertices and the transition rates $k_{ij}$ as weight for the directed edge $(i,j)$~\cite{chaiken1982combinatorial}.

If the network is irreducible, i.e., path-connected in the edge-directed sense, then there exists a unique stationary probability distribution $\pi$ satisfying $W\pi=0$~\cite{norris1998markov, seneta2006non}. The Markov chain tree theorem is a standard result stating that $\pi_n$ can be expressed as sums over spanning trees of the graph~\cite{hill1966studies,schnakenberg1976network,zia2007probability}:
\begin{align}
    \pi_n=\frac{\sum_{\mathcal{T}_{[n]}}w(\mathcal{T})}{\sum_{m=1}^N\sum_{\mathcal{T}_{[m]}}w(\mathcal{T})},\label{eq:MCTT}
\end{align}
where $\mathcal{T}_{[i]}$ denotes the spanning trees rooted at $i$, meaning that this vertex has out-degree 0 and the other vertices, $\{1,...,N\}\setminus \{i\}$, have out-degree $1$, and the weight $w(\mathcal{T})$ is defined as product of the edge weights over the edge set $E(\mathcal{T})$:
\begin{align}
    w(\mathcal{T}):=\prod_{(i,j)\in E(\mathcal{T})}k_{ij}.
\end{align}
Using this relation, Bebon and Speck~\cite{bebon2026mutual} recently derived an affine relation between the stationary occupation probabilities under variation of a single transition rate $k_{ij}$ termed mutual linearity:
\begin{align}
    \pi_n(k_{ij})=\pi_{n,m}^{(0)}+\chi_{n,m}^{(i,j)}\pi_m(k_{ij}),\label{eq:mutual_inearity_stationary}
\end{align}
where both $\pi_{n,m}^{(0)}$ and the susceptibility $\chi_{n,m}^{(i,j)}:=\frac{\partial_{k_{ij}}\pi_n}{\partial_{k_{ij}}\pi_m}$ are independent of $k_{ij}$. \new{Previously, Floyd et al.~\cite{floyd2025limits} had already derived a result in close relation to Eq.~(\ref{eq:mutual_inearity_stationary}) in the context of information processing in biochemical networks.} Earlier, Harunari et al.~\cite{harunari2024mutual} had established that a mutual linearity relation holds for stationary currents and for non-stationary (transient) currents in Laplace space. \new{The former result has also been proven by graph theoretic means~\cite{polettini2025coplanarity, khodabandehlou2025affine}.} Very recently, Zheng and Lu gave a trajectory-level derivation of mutual linearity based on the Doob-Meyer decomposition, showing that this property also extends to the non-stationary occupation probabilities in Laplace space \new{for Markovian jump and diffusion processes~\cite{zheng2026mutual,zheng2026mutual2}. Moreover, the stationary response relation for perturbing transition rates in opposite directions associated with mutual linearity also follows from a time-local factorization of the functional responses developed by Aslyamov and Esposito in the broader context of dynamical fluctuation-response relations~\cite{aslyamov2026dynamical}.} 

In this work, we derive an explicit graph-theoretic formulation of the resolvent $(s\mathbb{1}-W)^{-1}$, employing the all-minors matrix-tree theorem~\cite{chaiken1982combinatorial}, which allows for a combinatorial expression of the non-stationary occupation probabilities in Laplace space. This formulation yields an alternative proof of mutual linearity in Laplace space and explicit representations for physically relevant quantities such as the susceptibility in the non-stationary regime. We recover the mutual linearity relation in the stationary case as long-time limit.
Moreover, we show that the short-time scaling is determined by shortest path lengths from the initial distribution, \new{which is a consequence of the similar, well-established scaling of occupation probabilities for short times \cite{li2013mechanisms,li2014pathway}.} Because the
graph theoretic framework can also be used
to calculate the moments and the Laplace transform of the hitting time density \cite{nam2023linear,voits2024generic,nam2025algebraic, voits2026emergence}, we finally prove 
that a similar mutual linearity relation also holds between hitting time densities starting at different initial states.

\section{Combinatorial expression of the resolvent and the Laplace transformed probability distribution}

\begin{figure}
    \centering
    \includegraphics[width=\linewidth]{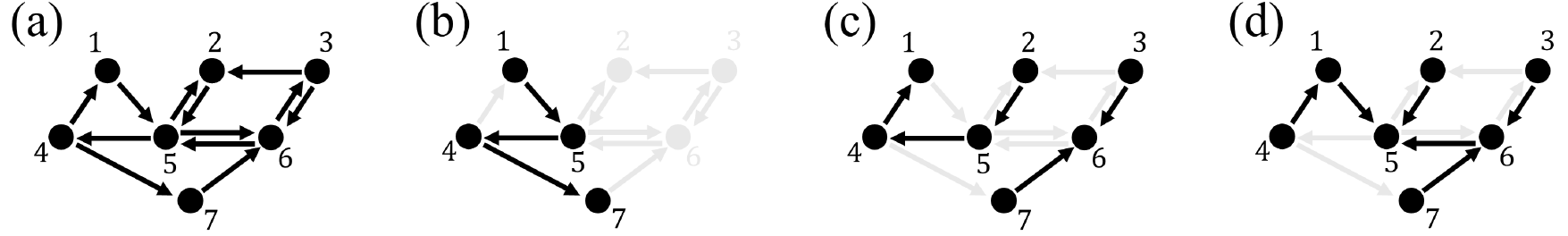}
    \caption{Illustration of relevant graph theoretic concepts. (a) An example of a directed graph $\mathcal{G}$ with $N=7$ vertices. (b) A path in  $\mathcal{G}$ from vertex $1$ to vertex $7$. (c) A spanning forest of $\mathcal{G}$ with roots $1$ and $6$. (d) A spanning tree of $\mathcal{G}$ with root $5$.}
    \label{fig:graph_theory_definitions}
\end{figure}

The combinatorial structure arises from describing the Markov network on $N$ states with jump rates $k_{ij}$ as a directed graph $\mathcal{G}=(V(\mathcal{G}),E(\mathcal{G}))$, with vertex set $V(\mathcal{G})=\{1,...,N\}$ and edge set $E(\mathcal{G})=\{(i,j)\in V(\mathcal{G})\times V(\mathcal{G})~|~k_{ij}>0\}$. As mentioned in the introduction, $-W$ is the out-degree
Laplacian matrix of that graph. 
An example network with $N=7$ states is shown in Fig.~\ref{fig:graph_theory_definitions}(a), with edge weights $k_{ij}$.
To make the combinatorial structure explicit, we will employ the following definitions and conventions:
\begin{itemize}
    \item A (non-self intersecting) path $\mathcal{P}^{m\to n}\subseteq \mathcal{G}$ from $m\in V(\mathcal{G})$ to $n\in V(\mathcal{G})$ with $V(\mathcal{P}^{m\to n})=\{m,i_1,i_2,...,i_{k-1},n\}$ has an edge set of the shape:
    \begin{align}
      E(\mathcal{P}^{m\to n}):=\{(m,i_1),(i_1,i_2),...(i_{k-2},i_{k-1}),(i_{k-1},n)\}\subseteq E(\mathcal{G}).
    \end{align}
    \new{The definition also permits} the trivial path $\mathcal{P}^{m\to m}=(\{m\},\emptyset)$\new{, which will be relevant for the proper definition of sets of subgraphs containing a path between two specific vertices}. An example for a path is shown in Fig.~\ref{fig:graph_theory_definitions}(b). 
    \item For a given set of roots $I=\{i_1,...,i_n\}\subseteq V(\mathcal{G})$, a spanning forest $\mathcal{F}_{[i_1,...i_n]}\subseteq\mathcal{G}$ is a subgraph of $\mathcal{G}$ with $V(\mathcal{F}_{[i_1,...i_n]})=V(\mathcal{G})$ such that for every $m\in V(\mathcal{F}_{[i_1,...i_n]})$, there exists exactly one path $\mathcal{P}^{m\to i}$ to exactly one root $i\in I$ (Fig.~\ref{fig:graph_theory_definitions}(c)). We denote the set of spanning forests with roots $I$ as $F_{{[i_1,...i_n]}}$. Moreover, a spanning forest with exactly one root, $n=1$, is called a tree (Fig.~\ref{fig:graph_theory_definitions}(d)). In this case, we write $\mathcal{T}_{[i]}$ and $T_{[i]}$, instead of $\mathcal{F}_{[i]}$ and $F_{[i]}$ to emphasize this structure.
    For a spanning forest $\mathcal{F}_{{[i_1,...i_n]}}$, the subgraph $\tau_{i_j}(\mathcal{F}_{{[i_1,...i_n]}})\subseteq \mathcal{F}_{{[i_1,...i_n]}}$ for $j\in\{1,...,n\}$ containing all the vertices with a path to $i_j$ and the corresponding edges, i.e.,
    \begin{align}
        V(\tau_{i_j})&=\{m\in V(\mathcal{F}_{{[i_1,...i_n]}})~|~\mathcal{F}_{{[i_1,...i_n]}}\text{ contains an $m\to i_j  $~-path}\}\\ E(\tau_{i_j})&=V(\tau_{i_j})\times V(\tau_{i_j})\cap E(\mathcal{F}_{{[i_1,...i_n]}}),
    \end{align}
    is referred to as the $i_j$-tree of $\mathcal{F}_{{[i_1,...i_n]}}$.
    \item We denote restrictions to a set of spanning forests as upper case indices. This includes restrictions to forests which contain a path between two specific vertices $m$ and $k$:
    \begin{align}
        F_{[i_1,...i_n]}^{m\to k}&:=F_{[i_1,...i_n]}\cap\{\mathcal{S}\subseteq\mathcal{G}~|~\mathcal{S}\text{ contains an $m\to k$-path}\},
    \end{align}
    and combinations thereof. 
\end{itemize}

The all-minors matrix-tree theorem gives a combinatorial expression for the determinants of the minors of $W$ in terms of sums over spanning forests \cite{chaiken1982combinatorial}. In particular, the following identity holds for the $k$-th principal minor:
\begin{align}
    \det (-W[\{k\}^c,\{k\}^c])=\sum_{\mathcal{T}_{[k]}}w(\mathcal{T}),\label{eq:principal_minor_identity}
\end{align}
where here and in the following the sum is to be understood as a short-hand notation for summing over all $\mathcal{T}_{[k]}\in T_{[k]}$.
Moreover, the following holds for minors of the shape $W[\{n,k\}^c,\{m,k\}^c]$, i.e., obtained by deleting the $n$-th and $k$-th rows and the $m$-th and $k$-th columns:
\begin{align}
    \det (-W[\{n,k\}^c,\{m,k\}^c])=(-1)^{n+m}\sum_{\mathcal{F}^{m\to n}_{[n,k]}}w(\mathcal{F}),\label{eq:non_principal_minor_identity}
\end{align}
where again, the sum is taken over all $\mathcal{F}^{m\to n}_{[n,k]}\in F^{m\to n}_{[n,k]}$.

Combining the two expressions \new{from Eq.~(\ref{eq:principal_minor_identity}) and Eq.~(\ref{eq:non_principal_minor_identity})}, the entries of the inverse of $- W[\{k\}^c,\{k\}^c]$ can be expressed as:
\begin{align}
    -W[\{k\}^c,\{k\}^c]^{-1}_{nm}&=(-1)^{n+m}\frac{\det (-W[\{n,k\}^c,\{m,k\}^c])}{\det(- W[\{k\}^c,\{k\}^c])}\\&=\frac{\sum_{\mathcal{F}^{m\to n}_{[n,k]}}w(\mathcal{F})}{\sum_{\mathcal{T}_{[k]}}w(\mathcal{T})},
\end{align}
showing explicitly the combinatorial structure of the inverse matrix.

\begin{figure}
    \centering
    \includegraphics[width=\linewidth]{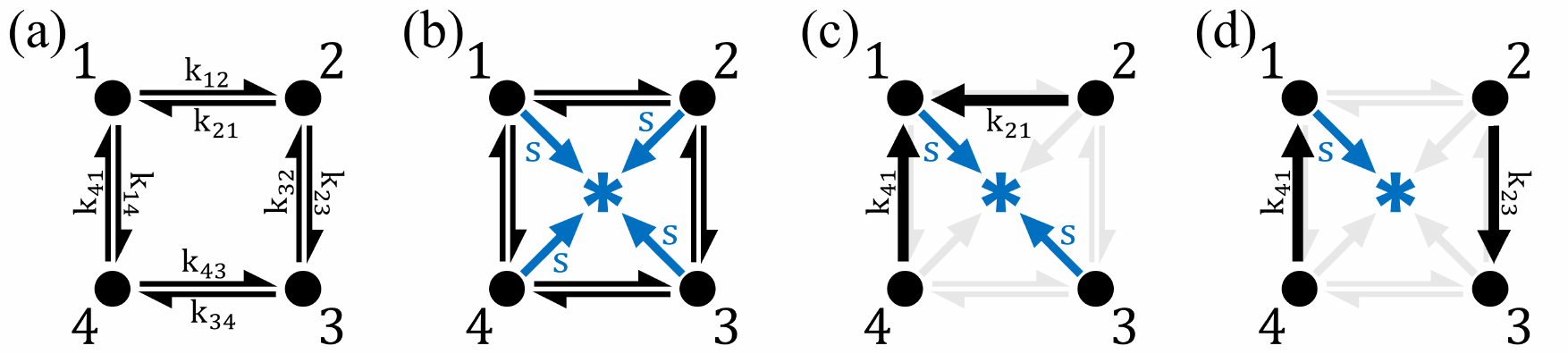}
    \caption{Illustration of the construction that allows expressing the Laplace transform of the occupation probabilities in graph theoretic terms. (a) An example for a Markov network on four states represented as a weighted and directed graph $\mathcal{G}$ with the states as vertices, the transitions as directed edges with the transition rates as edge weights. (b) The augmented graph $\mathcal{G}^\ast$ is constructed by including an additional vertex $\ast$ and edges with rate $s$ leading from every other vertex to $\ast$. (c) A spanning tree of the augmented graph with root in $\ast$. (d) A spanning forest containing two trees with roots in $3$ and $\ast$.}
    \label{fig:augmented_network}
\end{figure}

The Laplace transform,  defined for a function $g(t)$ as:
\begin{equation}
    \Laplace{g}=\integ{0}{\infty}{t} e^{-st}g(t),
\end{equation}
for $s\geq 0$, satisfies $\Laplace{\dot{g}}=s\Laplace{g}-g(0)$. Therefore, applying this transform converts Eq.~(\ref{eq:master_equation}) into an algebraic equation: 
\begin{align}
    s\Laplace{p}-p(0)&=W\Laplace{p} \\\Rightarrow (s\mathbb{1}-W)\Laplace{p}&=p(0)\label{eq:ME_Laplace_space},
\end{align}
where $\mathbb{1}$ denotes the identity matrix in $N\times N$ dimensions.
We note that $s\mathbb{1}-W$ can be interpreted as a principal minor of the out-degree Laplacian matrix of the graph $\mathcal{G}^\ast$ that is obtained by augmenting the vertex set of the original network by an additional state $\ast$, with incident edges with rate $s$ from any other vertex (Fig.~\ref{fig:augmented_network}(b)),
\begin{align}
    \mathcal{G}^\ast=\big(V(\mathcal{G})\cup\{*\},E(\mathcal{G})\cup\{(i,\ast)~|~i\in V(\mathcal{G})\}\big),~~k_{i\ast}:=s~~\forall i\in V(\mathcal{G}).
\end{align}

Hence, the entries of the resolvent $(s\mathbb{1}-W)^{-1}$ can be expressed as ratios of spanning forests and spanning trees of the augmented network: 
\begin{align}
    (s\mathbb{1}-W)^{-1}_{nm}=\frac{\sum_{\mathcal{F}^{s, m\to n}_{[n,*]}}w(\mathcal{F})}{\sum_{\mathcal{T}^{s}_{[*]}}w(\mathcal{T})}\label{eq:graph_theoretic_inverse}.
\end{align}
The denominator sums over spanning trees $\mathcal{T}^{s}_{[*]}$ with root $\ast$ of $\mathcal{G}^\ast$ (Fig.~\ref{fig:augmented_network}(c)), while the numerator sums over spanning forests $\mathcal{F}^{s,m\to n}_{[n,\ast]}$ consisting of two trees rooted at $n$ and $\ast$, with $m$ contained in the tree rooted at $n$ (Fig.~\ref{fig:augmented_network}(d)).
Thus, the Laplace transformed occupation probabilities can be expressed as:
\begin{align}
    \Laplace{p_n}&=\sum_{m=1}^N(s\mathbb{1}-W)^{-1}_{nm}~p_{m}(0)\\ &=\sum_{m=1}^N\frac{\sum_{\mathcal{F}^{s, m\to n}_{[n,*]}}w(\mathcal{F})}{\sum_{\mathcal{T}^{s}_{[*]}}w(\mathcal{T})}p_{m}(0)\label{eq:prob_density_Laplace_space}.
\end{align}

\section{Mutual linearity of the occupation probabilities in Laplace space}
\label{sec:prob_densities}

Mutual linearity of the probability distribution in Laplace space under variation of a transition rate $k_{ij}$ is the property that 
\begin{align}
    \{\mathcal{L}\{p\}(s,k_{ij})~|~k_{ij}>0\}\subseteq\mathbb{R}^N
\end{align}
lies on an affine subspace of $\mathbb{R}^N$. Equivalently, the Laplace transforms of two occupation probabilities $p_n(t)$ and $p_m(t)$, for $n,m\in\{1,...,N\}$ satisfy an affine relation (mutual linearity):
\begin{align}
    \Laplace{p_n}&=\Laplace{q_{n,m}^{(i,j)}}+\Laplace{\chi_{n,m}^{(i,j)}}\Laplace{p_m},\label{eq:mutual_linerarity_Laplace}
\end{align}
where both $\Laplace{q_{n,m}^{(i,j)}}$ and $\Laplace{\chi_{n,m}^{(i,j)}}$ are constant with respect to $k_{ij}$. This was shown by Zheng and Lu at the trajectory level based on the Doob-Meyer decomposition \cite{zheng2026mutual}. Here, we obtain \new{Eq.~(\ref{eq:mutual_linerarity_Laplace})} within the combinatorial framework. In addition, this approach results in explicit formulae for $\Laplace{p_m}$ and $\Laplace{\chi_{n,m}^{(i,j)}}$.
 
Taking the derivative of Eq.~(\ref{eq:ME_Laplace_space}) with respect to one rate $k_{ij}$ gives:
\begin{align}
   0&=-(\partial_{k_{ij}}W)\Laplace{p}+(s\mathbb{1}-W)\partial_{k_{ij}}\Laplace{p}\\\Rightarrow\partial_{k_{ij}}\Laplace{p}&=(s\mathbb{1}-W)^{-1}( \partial_{k_{ij}}W)\Laplace{p}.
\end{align}
Since the derivative of $W$ reads:
\begin{align}
    ( \partial_{k_{ij}}W)_{lm}=\delta_{lj}\delta_{mi}-\delta_{li}\delta_{lm},
\end{align}
applying the identity found in Eq.~(\ref{eq:graph_theoretic_inverse}) implies:
\begin{align}
   \partial_{k_{ij}}\Laplace{p_n}&=(s\mathbb{1}-W)^{-1}_{nl}( \partial_{k_{ij}}W)_{lm}\Laplace{p_m}\\ &=\frac{\sum_{\mathcal{F}^{s, j\to n}_{[n,*]}}w(\mathcal{F})-\sum_{\mathcal{F}^{s, i\to n}_{[n,*]}}w(\mathcal{F})}{\sum_{\mathcal{T}^{s}_{[*]}}w(\mathcal{T})}\Laplace{p_i}\label{eq:Laplace_partial_derivative_p_n}.
\end{align}
Therefore, it follows that the response ratios of the Laplace transform of the occupation probability of state $n$ under variation of a transition rate $k_{ij}$ and the reverse transition $k_{ji}$ is given by:
\begin{align}
   \frac{\partial_{k_{ij}}\Laplace{p_n}}{\partial_{k_{ji}}\Laplace{p_n}}&=-\frac{\Laplace{p_i}}{\Laplace{p_j}}\label{eq:ratio_k_ij_k_ji},
\end{align}
and the ratio for two different states $n$ and $m$ under variation of the same rate $k_{ij}$,
\begin{align}
    \Laplace{\chi_{n,m}^{(i,j)}}&:=\frac{\partial_{k_{ij}}\Laplace{p_n}}{\partial_{k_{ij}}\Laplace{p_m}} \\&= \frac{\sum_{\mathcal{F}^{s, j\to n}_{[n,*]}}w(\mathcal{F})-\sum_{\mathcal{F}^{s, i\to n}_{[n,*]}}w(\mathcal{F})}{\sum_{\mathcal{F}^{s, j\to m}_{[m,*]}}w(\mathcal{F})-\sum_{\mathcal{F}^{s, i\to m}_{[m,*]}}w(\mathcal{F})}\\&=\frac{\sum_{\mathcal{F}^{s, j\to n,i\to *}_{[n,*]}}w(\mathcal{F})-\sum_{\mathcal{F}^{s, i\to n, j\to *}_{[n,*]}}w(\mathcal{F})}{\sum_{\mathcal{F}^{s, j\to m,i\to*}_{[m,*]}}w(\mathcal{F})-\sum_{\mathcal{F}^{s, i\to m, j\to *}_{[m,*]}}w(\mathcal{F})}\label{eq:def_chi},
\end{align}
\new{where the last line follows from observing that all spanning forests of the shape $\mathcal{F}^{s, i\to n,j\to n}_{[n,*]}$ and $\mathcal{F}^{s, i\to m,j\to m}_{[m,*]}$ appear in both sums in the numerator and the denominator, respectively.}
Thus, \new{Eq.~(\ref{eq:def_chi}) provides an} explicit combinatorial expression for the Laplace transform of the susceptibility $\chi_{n,m}^{(i,j)}(t)$ and the following properties are immediate corollaries of it:
\begin{align}
    \Laplace{\chi_{n,m}^{(i,j)}}&=\Big(\Laplace{\chi_{m,n}^{(i,j)}}\Big)^{-1},\\\Laplace{\chi_{n,m}^{(i,j)}}&=\Laplace{\chi_{n,m}^{(j,i)}},\label{eq:exchange_symmetry_chi}\\\Laplace{\chi_{n,m}^{(i,j)}}&=\Laplace{\chi_{n,l}^{(i,j)}}\Laplace{\chi_{l,m}^{(i,j)}}.
\end{align}
Also, $\chi_{n,m}^{(i,j)}(t)$ does not depend on the initial distribution $p(0)$.
Moreover, note that the sums in both the numerator and the denominator of Eq.~(\ref{eq:def_chi}) are constant with respect to $k_{ij}$, so
\begin{align}
    \partial_{k_{ij}}\sum_{\mathcal{F}^{s, j\to l,i\to *}_{[l,*]}}w(\mathcal{F})=\partial_{k_{ij}}\sum_{\mathcal{F}^{s, i\to l, j\to *}_{[l,*]}}w(\mathcal{F})=0,
\end{align}
for all $l\in V(\mathcal{G})$, since the edge $(i,j)$ can only be contained in forests where $i$ and $j$ belong to the same tree.
Hence, the susceptibility is also constant with respect to $k_{ij}$,
\begin{align}
    \partial_{k_{ij}}\Laplace{\chi_{n,m}^{(i,j)}}=0,
\end{align}
which implies that
\begin{align}
    \partial_{k_{ij}}\Laplace{p_n}&=\Laplace{\chi_{n,m}^{(i,j)}}\partial_{k_{ij}}\Laplace{p_m}\\ &=\partial_{k_{ij}}\big(\Laplace{\chi_{n,m}^{(i,j)}}\Laplace{p_m}\big)\\\Rightarrow \Laplace{p_n}&=\Laplace{q_{n,m}^{(i,j)}}+\Laplace{\chi_{n,m}^{(i,j)}}\Laplace{p_m},\label{eq:MutLin_Laplace}
\end{align}
where $\Laplace{q_{n,m}^{(i,j)}}$ does not depend on $k_{ij}$. This yields the mutual linearity relation of the probability distribution in Laplace space that was shown in~\cite{zheng2026mutual}.
By the convolution theorem for the Laplace transform, the product of two Laplace-transformed functions corresponds to their convolution in the time domain and therefore, 
\begin{align}
    p_n(t)&=q_{n,m}^{(i,j)}(t)+(\chi_{n,m}^{(i,j)}\ast p_m)(t)\\ &\new{=q_{n,m}^{(i,j)}(t)+\integ{0}{t}{t'}\chi_{n,m}^{(i,j)}(t-t') p_m(t')}.\label{eq:MutLin_real}
\end{align}

\begin{figure*}[h]
    \centering
    \includegraphics[width=0.8\linewidth]{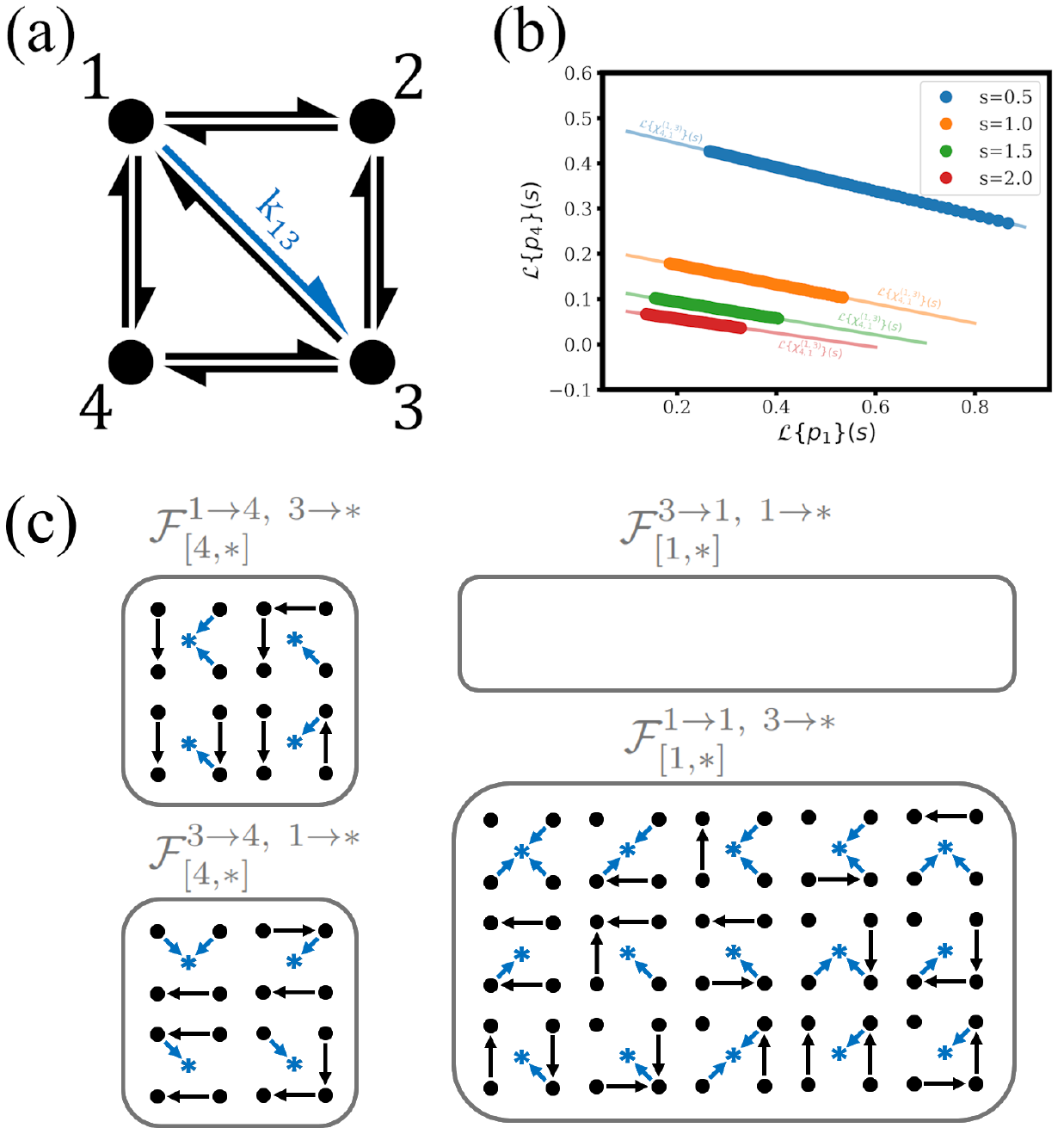}
    \caption{Illustration of the mutual linearity for an example network \new{and the underlying combinatorial structure}. (a) Network with four vertices, where the rate $k_{13}$ is varied. (b) Affine relation (mutual linearity) between the Laplace transforms of the occupation probabilities $\Laplace{p_1}$ and $\Laplace{p_2}$ for different values of $s$. 
    The parameters are $k_{12}=1$, $k_{21}=0.4$, $k_{23}=1.2$, $k_{32}=0.5$, $k_{34}=0.9$, $k_{43}=0.6$,  $k_{41}=1.1$, $k_{14}=0.3$, $k_{31}=0.2$ and $k_{13}\in [0,5]$. The solution is obtained from a numerical inversion of Eq.~(\ref{eq:ME_Laplace_space}). The transparent lines show the analytical solution for the slope, \new{$\mathcal{L} \{\chi_{4,1}^{(1,3)}\}(s)=\frac{k_{14}(s+k_{21}+k_{23}+k_{32})-k_{34}(s+k_{12}+k_{21}+k_{23})}{
    (s+k_{21}+k_{23})(s+k_{34}+k_{41}+k_{43})+k_{32}(s+k_{41}+k_{43})}$, obtained from the graph theoretic expression, Eq.~(\ref{eq:def_chi}). (c) Two-tree spanning forests of the augmented graph that appear in the combinatorial expression for $\mathcal{L} \{\chi_{4,1}^{(1,3)}\}(s)$. For the two sums in the numerator of Eq.~(\ref{eq:def_chi}), there are four forests each. There are no combinations for the first sum in the denominator, but fifteen two-tree spanning forests for the second sum. The $s$-rates are highlighted in blue.}}
    \label{fig:mutual_linearity_illustration}
\end{figure*}

Note that Eq.~(\ref{eq:MutLin_Laplace}) has the geometric interpretation that, for any fixed $s$, varying the rate $k_{ij}$ moves the vector $\Laplace{p}$ along a one-dimensional affine subspace of $\mathbb{R}^N$, as illustrated in Fig.~\ref{fig:mutual_linearity_illustration} for an example network. \new{Importantly, Eq.~(\ref{eq:MutLin_Laplace}) therefore corresponds to a convolution relation in the time domain as shown in Eq.~(\ref{eq:MutLin_real}) rather than to an affine relation between $p_n(t)$ and $p_m(t)$ at a fixed time $t$. Instead, $p_n(t)$ depends on the history of $p_m(t)$ through $\chi_{n,m}^{(i,j)}(t)$, which plays the role of a memory kernel.}

Following the analysis of Bebon and Speck for the stationary distribution~\cite{bebon2026mutual}, we introduce the limit $k_{ij}\to\infty$ of $\Laplace{p_n}$ and use the combinatorial representation Eq.~(\ref{eq:prob_density_Laplace_space}) to evaluate it:
\begin{align}
    \mathcal{B}_n^{(i,j)}(s)&:=\lim_{k_{ij}\to\infty}\Laplace{p_n}\\ &=\lim_{k_{ij}\to\infty}\sum_{m=1}^N\frac{\sum_{\mathcal{F}^{s, m\to n}_{[n,*]}}w(\mathcal{F})}{\sum_{\mathcal{T}^{s}_{[*]}}w(\mathcal{T})}p_{m}(0)\\ &= \sum_{m=1}^N\frac{\sum_{\mathcal{F}^{s, m\to i, j\to n}_{[n,i,*]}}w(\mathcal{F})+\sum_{\mathcal{F}^{s, m\to n,j\to n}_{[n,i,*]}}w(\mathcal{F})+\sum_{\mathcal{F}^{s, m\to n,j\to *}_{[n,i,*]}}w(\mathcal{F})}{\sum_{\mathcal{F}^{s, j\to *}_{[i,*]}}w(\mathcal{F})}p_{m}(0),\label{eq:def_Bn}
\end{align}
as only forests containing the edge $(i,j)$ remain relevant in this limit. In particular, the limit exists. By Eq.~(\ref{eq:exchange_symmetry_chi}), exchanging $i$ and $j$ leaves the susceptibility unchanged, so varying $k_{ij}$ or $k_{ji}$ generates the same affine subspace.
Since the two points
\begin{align}
    \lim_{k_{ij}\to\infty}\big(\Laplace{p_m},\Laplace{p_n}\big)&=\big(\mathcal{B}_m^{(i,j)}(s),\mathcal{B}_n^{(i,j)}(s)\big),\\\lim_{k_{ji}\to\infty}\big(\Laplace{p_m},\Laplace{p_n}\big)&=\big(\mathcal{B}_m^{(j,i)}(s),\mathcal{B}_n^{(j,i)}(s)\big)
\end{align}
lie on the affine subspace containing $\big(\mathcal{L}\{p\}(s,k_{ij})\big)_{k_{ij}>0}$, an explicit expression for $\Laplace{q_{n,m}^{(i,j)}}$ is given by
\begin{align}
    \Laplace{q_{n,m}^{(i,j)}}=\frac{\mathcal{B}_n^{(j,i)}(s)\mathcal{B}_m^{(i,j)}(s)-\mathcal{B}_m^{(j,i)}(s)\mathcal{B}_n^{(i,j)}(s)}{\mathcal{B}_m^{(i,j)}(s)-\mathcal{B}_m^{(j,i)}(s)}.
\end{align}

\section{Long-time limits of mutual linearity of the probability distribution}
\label{sec:asymptotics1}

The explicit combinatorial expressions given in the previous section allow us to recover the results for the stationary limit ($t\to\infty$ or long-time limit). In particular, we show that 
our results are equivalent to the ones obtained by Bebon and Speck~\cite{bebon2026mutual}.
The long-time limit corresponds to the limit $s\to 0$ in Laplace space. \new{Since all non-zero eigenvalues of $W$ have negative real part, we have 
\begin{align}
    \det(s\mathbb{1}-W)=\sum_{\mathcal{T}^{s}_{[*]}}w(\mathcal{T})\neq0,
\end{align}
for $s\in \mathbb{C}\setminus\{0\}$ with $ \text{Re}(s)\geq 0$}. Therefore, Eqs.~(\ref{eq:prob_density_Laplace_space}) and (\ref{eq:Laplace_partial_derivative_p_n}) imply that both $\Laplace{p_n}$ and $\partial_{k_{ij}}\Laplace{p_n}$ and\new{, by extension, $\Laplace{\chi_{n,m}^{(i,j)}}$ have no non-zero poles with $\text{Re}(s)\geq 0$}. \new{Moreover, if the graph is strongly connected, the pole at $s=0$ is simple.}  Thus, the \new{final}-value theorem for the Laplace transform~\cite{doetsch2012introduction} can be applied on these functions, which states that, \new{if all poles of $s\mathcal{L}{f}(s)$ lie in the left half-plane, except possibly a simple pole at $s=0$,} then the limits in the time domain and Laplace space \new{of $f$} are related as
\begin{align}
    \lim_{t\to\infty}f(t)=\lim_{s\to0}s\Laplace{f}.
\end{align}
Note that for an irreducible network, i.e., for a strongly connected graph $\mathcal{G}$, this limit relation implies:
\begin{align}
    \lim_{t\to \infty}p_n(t)&=\lim_{s\to0}s\Laplace{p_n}\\ &=\sum_{m=1}^Np_{m}(0)\lim_{s\to0}\frac{s\sum_{\mathcal{F}^{s, m\to n}_{[n,*]}}w(\mathcal{F})}{\sum_{\mathcal{T}^{s}_{[*]}}w(\mathcal{T})}\\ &=\sum_{m=1}^Np_{m}(0)\frac{\sum_{\mathcal{T}_{[n]}}w(\mathcal{T})}{\sum_{l=1}^N\sum_{\mathcal{T}_{[l]}}w(\mathcal{T})}\\ &=\frac{\sum_{\mathcal{T}_{[n]}}w(\mathcal{T})}{\sum_{l=1}^N\sum_{\mathcal{T}_{[l]}}w(\mathcal{T})},
\end{align}
recovering the identity in Eq.~(\ref{eq:MCTT}), with $\lim_{t\to \infty}p_n(t)=\pi_n$. The last step follows from the normalization of the probabilities, and the limit $s\to 0$ is determined by the lowest order coefficients in the numerator and the denominator. For the spanning forests $\mathcal{F}^{s, m\to n}_{[n,*]}$ of $\mathcal{G}^\ast$, a constant contribution comes from forests where $\ast$ is an isolated vertex, so all other vertices are contained in the tree with root $n$, so their weights correspond to the weights of the spanning trees of $\mathcal{G}$ with root $n$. In the denominator, the lowest order is linear in $s$ and comes from the  
spanning trees $\mathcal{T}^{s}_{[*]}$ with exactly one edge $(l,\ast)$ for an $l\in\mathbb{N}$. Thus, the coefficient is the sum of the weights of all spanning trees of $\mathcal{G}$.
For the stationary limit of Eq.~(\ref{eq:ratio_k_ij_k_ji}), we then obtain:
\begin{align}
     \frac{\partial_{k_{ij}}\pi_n}{\partial_{k_{ji}}\pi_n}&=\lim_{s\to 0}\frac{\partial_{k_{ij}}\big(s\Laplace{p_n^s}\big)}{\partial_{k_{ji}}\big(s\Laplace{p_n^s}\big)}\\ &=\lim_{s\to 0}\frac{\partial_{k_{ij}}\Laplace{p_n^s}}{\partial_{k_{ji}}\Laplace{p_n^s}}\\ &=-\lim_{s\to 0}\frac{\Laplace{p_i}}{\Laplace{p_j}}\\&=-\frac{\pi_i}{\pi_j}.\label{eq:stationary_reverse_edge_response}
\end{align}

Similarly, by considering the lowest order terms in $s$, the limits of Eqs.~(\ref{eq:def_Bn}) and (\ref{eq:def_chi}) can be computed:
\begin{align}
   \mathcal{B}_n^{s,(i,j)}&=\lim_{s\to0}s\new{\mathcal{B}_n^{(i,j)}(s)}\\&=\frac{\sum_{\mathcal{F}^{j\to n}_{[i,n]}}w(\mathcal{F})}{\sum_{l}\sum_{\mathcal{F}^{j\to l}_{[i,l]}}w(\mathcal{F})}\\&=\frac{\sum_{\mathcal{T}_{[n]},(i,j)\in E(\mathcal{T})}w(\mathcal{T})}{\sum_{\mathcal{T},{(i,j)\in E(\mathcal{T})}}w(\mathcal{T})},\label{eq:Bnsji}\\ \chi_{n,m}^{s,(i,j)}&=\lim_{s\to0}\new{\Laplace{\chi_{n,m}^{(i,j)}}}\\&=\frac{\sum_l\sum_{\mathcal{F}^{\tiny\substack{j\to n\\i\to l}}_{[n,l]}}w(\mathcal{F})-\sum_{\mathcal{F}^{\tiny\substack{i\to n\\j\to l}}_{[n,l]}}w(\mathcal{F})}{\sum_l\sum_{\mathcal{F}^{ \tiny\substack{j\to m\\ i\to l}}_{[m,l]}}w(\mathcal{F})-\sum_{\mathcal{F}^{ \tiny\substack{i\to m\\j\to l}}_{[m,l]}}w(\mathcal{F})},\label{eq:stationary_chi}
\end{align}
where the \new{superscript} $s$ denotes the stationary case.
While the expression for $\mathcal{B}_n^{s,(i,j)}$ (Eq.~(\ref{eq:Bnsji})) is the same as the stationary solution given by Bebon and Speck~\cite{bebon2026mutual}, their result for the susceptibility in the stationary case reads:
\begin{align}
    \chi_{n,m}^{s,(i,j)}&=\frac{\mathcal{B}_n^{s,(i,j)}-\mathcal{B}_n^{s,(j,i)}}{\mathcal{B}_m^{s,(i,j)}-\mathcal{B}_m^{s,(j,i)}}\label{eq:chi_Bebon_Speck}.
\end{align}
To see that the two expressions, \new{Eqs.~(\ref{eq:stationary_chi}) and (\ref{eq:chi_Bebon_Speck})}, are indeed equivalent, note that
\begin{align}
    \mathcal{B}_n^{s,(i,j)}-\mathcal{B}_n^{s,(j,i)}&=\frac{\sum_{l}\sum_{\mathcal{F}^{i\to l}_{[j,l]}}w(\mathcal{F})\sum_{\mathcal{F}^{j\to n}_{[i,n]}}w(\mathcal{F})-\sum_{l}\sum_{\mathcal{F}^{j\to l}_{[i,l]}}w(\mathcal{F})\sum_{\mathcal{F}^{i\to n}_{[j,n]}}w(\mathcal{F})}{\sum_{l}\sum_{\mathcal{F}^{j\to l}_{[i,l]}}w(\mathcal{F})\sum_{l}\sum_{\mathcal{F}^{i\to l}_{[j,l]}}w(\mathcal{F})}\label{eq:Bij-Bji}
\end{align}
contains terms in the numerator that appear in both products and cancel due to the difference. These are precisely the terms where the $i\to l$-path in $\mathcal{F}^{i\to l}_{[j,l]}$ intersects $\tau_n(\mathcal{F}^{j\to n}_{[i,n]})$, the $n$-tree of $\mathcal{F}^{j\to n}_{[i,n]}$ and where the $j\to n$-path in $\mathcal{F}^{j\to n}_{[i,n]}$ intersects the $l$-tree of $\mathcal{F}^{i\to l}_{[j,l]}$, $\tau_l(\mathcal{F}^{i\to l}_{[j,l]})$. 

\begin{figure}[t]
    \centering
    \includegraphics[width=\linewidth]{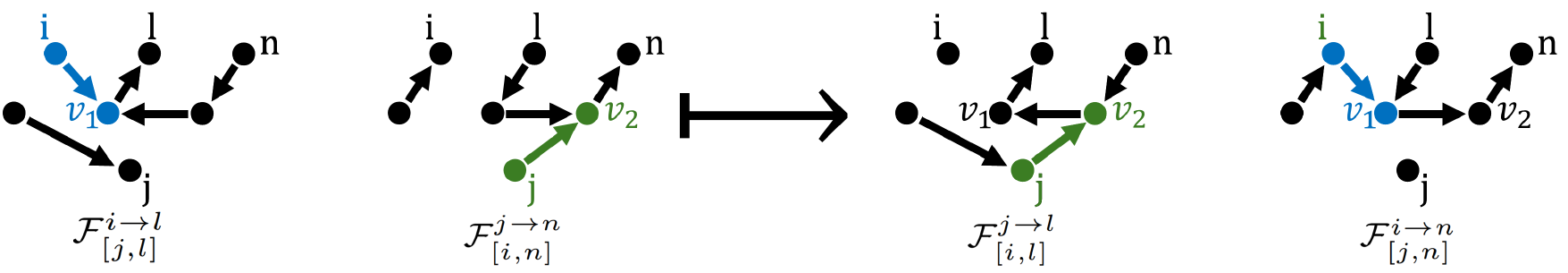}
    \caption{Illustration of the auxiliary map $H^{ij}_{nl}$ which exchanges the outgoing edges of the vertices on the $i\to l$-path in $\mathcal{F}^{i\to l}_{[j,l]}$ until it reaches the $n$-tree $\mathcal{F}^{j\to n}_{[i,n]}$ (blue path) with the corresponding edges in $\mathcal{F}^{j\to n}_{[i,n]}$, and also exchanges the outgoing edges of the vertices on the $j\to n$-path in $\mathcal{F}^{j\to n}_{[i,n]}$ until it reaches the $l$-tree $\mathcal{F}^{i\to l}_{[j,l]}$ (green path) with the corresponding edges in $\mathcal{F}^{i\to l}_{[j,l]}$. Thereby, $H^{ij}_{nl}$ yields a tuple $(\mathcal{F}^{j\to l}_{[i,l]},\mathcal{F}^{i\to n}_{[j,n]})$.}
    \label{fig:H_map}
\end{figure}

To see this, we show that these can be mapped onto each other. \new{This proof strategy is similar to the method of 'tree surgery'~\cite{owen2020universal,polettini2025coplanarity} but applied here to two-tree spanning forests.} First, we define for $S \subseteq F\times F$, where $F$ is the set of all spanning forests of $\mathcal{G}$,
\begin{align}
    S\Big|^{\beta_1\overset{\tau_{\alpha_1}}{\to} \gamma_1}_{\beta_2\overset{\tau_{\alpha_2}}{\to} \gamma_2}:=&\{(\mathcal{F}
    _1,\mathcal{F}_2)\in S ~|~\beta_1\to\gamma_1\text{-path }\mathcal{P}^{\beta_1\to \gamma_1}\subseteq \mathcal{F}_2,V(\tau_{\alpha_1}(\mathcal{F}_1))\cap V(\mathcal{P}^{\beta_1\to \gamma_1})\neq\emptyset\}\\\cap&\{(\mathcal{F}
    _1,\mathcal{F}_2)\in S ~|~\beta_2\to\gamma_2\text{-path }\mathcal{P}^{\beta_2\to \gamma_2}\subseteq \mathcal{F}_1,V(\tau_{\alpha_2}(\mathcal{F}_2))\cap V(\mathcal{P}^{\beta_2\to \gamma_2}) \neq\emptyset\}.
\end{align}
 Now, consider the auxiliary map:
\begin{align}
    H^{ij}_{nl}:F^{i\to l}_{[j,l]}\times F^{j\to n}_{[i,n]}\Big|^{i\overset{\tau_n}{\to} l}_{j\overset{\tau_l}{\to} n}&\to F^{j\to l}_{[i,l]}\times F^{i\to n}_{[j,n]}\Big|^{j\overset{\tau_n}{\to} l}_{i\overset{\tau_l}{\to} n}\\ (\mathcal{F}^{i\to l}_{[j,l]},\mathcal{F}^{j\to n}_{[i,n]})&\mapsto (\mathcal{F}^{j\to l}_{[i,l]},\mathcal{F}^{i\to n}_{[j,n]}),
\end{align}
defined by considering the $i\to l$-path $\mathcal{P}^{i\to l}\subseteq \mathcal{F}^{i\to l}_{[j,l]}$ up to the last vertex $v_1$ until it first intersects $\tau_n(\mathcal{F}^{j\to n}_{[i,n]})$, to obtain $i\to v_1$ and $v_1\to l$ paths $\mathcal{P}^{i\to v_1},\mathcal{P}^{v_1\to l}\subseteq \mathcal{F}^{i\to l}_{[j,l]}$ and similarly, choosing vertex $v_2$ as the first vertex where the $j\to n$ path $\mathcal{P}^{j\to n}\subseteq\mathcal{F}^{j\to n}_{[i,n]}$ intersects $\tau_l(\mathcal{F}^{i\to l}_{[j,l]})$ for the first time to construct $j\to v_2$ and $v_2\to n$ paths, $\mathcal{P}^{j\to v_2}, \mathcal{P}^{v_2\to n}\subseteq\mathcal{F}^{j\to n}_{[i,n]}$. By construction, $E(\mathcal{P}^{i\to v_1})$ is disjoint from $E(\mathcal{P}^{j\to n})=E(\mathcal{P}^{j\to v_2})\cup E(\mathcal{P}^{v_2\to n})$, and $E(\mathcal{P}^{j\to v_2})$ is disjoint from $E(\mathcal{P}^{i\to l})=E(\mathcal{P}^{i\to v_1})\cup E(\mathcal{P}^{v_1\to l})$. $H^{ij}_{nl}$ then exchanges the edges of $\mathcal{P}^{i\to v_1}$ in $\mathcal{F}^{i\to l}_{[j,l]}$ with the outgoing edges of the same vertices in $\mathcal{F}^{j\to n}_{[i,n]}$.  Similarly, \new{$H^{ij}_{nl}$ exchanges} the edges of $\mathcal{P}^{j\to v_2}$ in $\mathcal{F}^{j\to n}_{[i,n]}$ with the outgoing edges of the same vertices in $\mathcal{F}^{i\to l}_{[j,l]}$. This preserves the forest structure but exchanges the roots $i$ and $j$, and also the intersection property, since $\mathcal{F}^{j\to l}_{[i,l]}$ now contains a $j\to l$ path via $v_2$, while the $\mathcal{P}^{v_2\to n}$ path is preserved in $\mathcal{F}^{i\to n}_{[j,n]}$, and $\mathcal{F}^{i\to n}_{[j,n]}$ now contains an $i\to n$ path via $v_1$, while the $\mathcal{P}^{v_1\to l}$ path is preserved in $\mathcal{F}^{j\to l}_{[i,l]}$. The mapping is illustrated in Fig.~\ref{fig:H_map}.
Note that $H^{ij}_{nl}$ is bijective with the inverse given as $(H^{ij}_{nl})^{-1}=H^{ji}_{nl}$.

\begin{figure}[t]
    \centering
    \includegraphics[width=\linewidth]{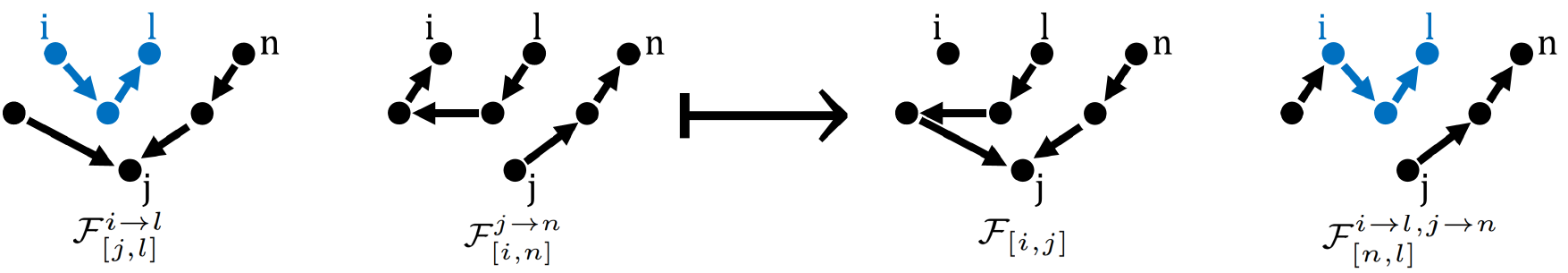}
    \caption{Illustration of the auxiliary map $A^{ij}_{nl}$ which exchanges the edges on the $i\to l$-path in $\mathcal{F}^{i\to l}_{[j,l]}$ with the corresponding edges in $\mathcal{F}^{j\to n}_{[i,n]}$ or on the $j\to n$-path in $\mathcal{F}^{j\to n}_{[i,n]}$ with the corresponding edges in $\mathcal{F}^{i\to l}_{[j,l]}$, yielding a tuple $(\mathcal{F}_{[i,j]},\mathcal{F}^{i\to l,j\to n}_{[n,l]})$.}
    \label{fig:H_map_2}
\end{figure}

The remaining terms, i.e., the forest tuples $\big(\mathcal{F}^{i\to l}_{[j,l]},\mathcal{F}^{j\to n}_{[i,n]}\big)\in F^{i\to l}_{[j,l]}\times F^{j\to n}_{[i,n]}$, where all vertices of the $i\to l$-path in the first graph are entirely contained in $\tau_i(\mathcal{F}^{j\to n}_{[i,n]})$ or where the vertices of the $j\to n$-path are entirely contained in $\tau_j(\mathcal{F}^{i\to l}_{[j,l]})$, can be bijectively mapped to $F^{i\to l, j\to n}_{[l,n]}\times F_{[i,j]}$, which can be seen by considering the following auxiliary map:
\begin{align}
    A^{ij}_{nl}:F^{i\to l}_{[j,l]}\times F^{j\to n}_{[i,n]}\setminus F^{i\to l}_{[j,l]}\times F^{j\to n}_{[i,n]}\Big|_{j\overset{\tau_l}{\to} n}^{i\overset{\tau_n}{\to} l}&\to F_{[i,j]} \times F^{i\to l, j\to n}_{[l,n]} \\ (\mathcal{F}^{i\to l}_{[j,l]},\mathcal{F}^{j\to n}_{[i,n]})&\mapsto (\mathcal{F}_{[i,j]},\mathcal{F}^{i\to l,j\to n}_{[l,n]}),
\end{align}
which is defined by: if the $i\to l$-path $\mathcal{P}^{i\to l}\subseteq\mathcal{F}^{i\to l}_{[j,l]}$ contains only vertices in 
$\tau_i(\mathcal{F}^{j\to n}_{[i,n]})$, exchange its edges with the corresponding edges in $\mathcal{F}^{j\to n}_{[i,n]}$, which turns the first forest into a forest with roots $i$ and $j$ and the second forest has roots $l$ and $n$ as well as $i\to l$- and $j\to n$-paths. Otherwise, the $j\to n$-path can be exchanged with the edges in $F^{i\to l}_{[j,l]}$ to obtain an analogous result. This map is illustrated in Fig.~\ref{fig:H_map_2}. Moreover, the mapping can be inverted by choosing the $i\to l$-path $\mathcal{F}^{i\to l,j\to n}_{[n,l]}$ and exchanging its edges with the corresponding edges in $\mathcal{F}_{[i,j]}$, which in turn yields a tuple in 
\begin{align}
    F^{i\to l}_{[j,l]}\times F^{j\to n}_{[i,n]}\setminus F^{i\to l}_{[j,l]}\times F^{j\to n}_{[i,n]}\Big|_{j\overset{\tau_l}{\to} n}^{i\overset{\tau_n}{\to} l}.
\end{align}

Thus, Eq.~(\ref{eq:chi_Bebon_Speck}) can be written as
\begin{align}
    \chi_{n,m}^{s,(i,j)}&=\frac{\mathcal{B}_n^{s,(i,j)}-\mathcal{B}_n^{s,(j,i)}}{\mathcal{B}_m^{s,(i,j)}-\mathcal{B}_m^{s,(j,i)}}\\&=\frac{\big(\sum_l\sum_{\mathcal{F}^{j\to n,i\to l}_{[n,l]}}w(\mathcal{F})-\sum_{\mathcal{F}^{i\to n,j\to l}_{[n,l]}}w(\mathcal{F})\big)\sum_{\mathcal{F}_{[i,j]}}w(\mathcal{F})}{\big(\sum_l\sum_{\mathcal{F}^{ j\to m, i\to l}_{[m,l]}}w(\mathcal{F})-\sum_{\mathcal{F}^{ i\to m,j\to l}_{[m,l]}}w(\mathcal{F})\big) \sum_{\mathcal{F}_{[i,j]}}w(\mathcal{F})}\\&=\frac{\sum_l\sum_{\mathcal{F}^{j\to n,i\to l}_{[n,l]}}w(\mathcal{F})-\sum_{\mathcal{F}^{i\to n,j\to l}_{[n,l]}}w(\mathcal{F})}{\sum_l\sum_{\mathcal{F}^{ j\to m, i\to l}_{[m,l]}}w(\mathcal{F})-\sum_{\mathcal{F}^{ i\to m,j\to l}_{[m,l]}}w(\mathcal{F})},
\end{align}
showing that this expression is equivalent to the result in Eq.~(\ref{eq:stationary_chi}).

\section{Short-time limits of mutual linearity of the probability distribution}
\label{sec:asymptotics2}

\new{It is a standard result that the short-time scaling of the occupation probabilities can be obtained from an expansion of the formal solution of the master equation to the first non-vanishing order~\cite{van1983stochastic,norris1998markov}. Defining the graph distance between $m$ and $n$ as
\begin{align}
    d(m,n):=\min_{\mathcal{P}^{m\to n}}|E(\mathcal{P}^{m\to n})|,
\end{align} 
such an expansion yields for the initial condition $p_n(0)=\delta_{nm}$:
\begin{align}
    p_n(t&)=(e^{Wt})_{nm}\\ &=\sum_{l=0}^\infty \frac{t^l}{l!}(W^{l})_{nm}\\ &=\frac{t^{d(m,n)}}{d(m,n)!}\sum_{\substack{\mathcal{P}^{m\to n}\\|E(\mathcal{P})|=d(m,n)}}w(\mathcal{P})+\mathcal{O}(t^{d(m,n)+1}),\label{eq:occupation_prob_short_times}
\end{align}
using that $(W^{l})_{nm}=0$ for $l<d(m,n)$ and for $l=d(m,n)$, it yields the sum of the weights of the minimal paths from $m\to n$. Physically, Eq.~(\ref{eq:occupation_prob_short_times}) means that at least $d(m,n)$ jumps have to happen in order to reach $n$ from $m$, and since for short times $\Delta t$, the probability for a jump from $i$ to $j$ is proportional to $k_{ij}\Delta t$, this yields the dependence on the shortest paths as given above.

This short-time scaling can also be derived from the large-$s$ behavior of Eq.~(\ref{eq:prob_density_Laplace_space}).} In this limit, the combinatorial terms are dominated by the trees and forests of $\mathcal{G}^\ast$ with the most edges to $\ast$. For $\sum_{\mathcal{T}^{s}_{[*]}}w(\mathcal{T})$, this is the tree with edges $\{(i,\ast)~|~i\in\{1,...,N\}\}$, which has the weight $s^N$. Note that the forests $\mathcal{F}^{s, m\to n}_{[n,\ast]}$ of highest order in $s$ are the forests that consist of a minimal path $\mathcal{P}^{m\to n}$ such that $|E(\mathcal{P}^{m\to n})|=d(m,n)$, and the weight of these forests is proportional to $s^{N-1-d(m,n)}$.

Hence, Eq.~(\ref{eq:prob_density_Laplace_space}) implies that the leading contribution to $\Laplace {p_n}$ arises from the minimal paths connecting states with positive occupation probability to $n$:
\begin{align}
    \Laplace{p_n}\overset{s\to\infty}{\simeq}&\sum_{\substack{m\\d(m,n)=d(p(0),n)}}p_m(0) \sum_{\substack{\mathcal{P}^{m\to n}\\|E(\mathcal{P})|=d(m,n)}}w(\mathcal{P})s^{-d(p(0),n)-1},\label{eq:short_time_scaling_Laplace_space}
\end{align}
where the graph distance
\begin{align}
    d(p(0),n):=\min_{\{l\in\{1,...,N\}~|~p_l(0)>0\}}d(l,n)
\end{align}
denotes the minimal number of transitions required to reach state $n$ from the initially occupied states $p_{l}(0)>0$, so that the inner sum runs over all directed paths $\mathcal P^{m\to n}$ of minimal length connecting $m$ to $n$.

\new{Eq.~(\ref{eq:short_time_scaling_Laplace_space})} also implies that the probability of occupying state $n$ vanishes to order $d(p(0),n)-1$ at short times,
\begin{align}
    p_n(0)=p_n^{(1)}(0)=\dots=p_n^{(d(p(0),n)-1)}(0)=0,\label{eq:derivatives_vanish}
\end{align}
reflecting that at least $d(p(0),n)$ transitions are required to reach $n$.

Applying the short-time expansion to Eq.~(\ref{eq:ratio_k_ij_k_ji}) yields:
\begin{align}
    \frac{\partial_{k_{ij}}\Laplace {p_n}}{\partial_{k_{ji}}\Laplace {p_n}}\overset{s\to\infty}{\simeq}- C_{ij}\, s^{\Delta_{ij}},
\end{align}
where $\Delta_{ij}:= d(p(0),j)-d(p(0),i)$, and
\begin{align}
    C_{ij}:=&\frac{\sum_{\substack{m:~d(m,i)=d(p_{0},i)}}p_{0,m} \sum_{\substack{\mathcal{P}^{m\to i}\\|E(\mathcal{P})|=d(m,i)}}w(\mathcal{P})}{\sum_{\substack{m:~d(m,j)=d(p_{0},j)}}p_{0,m} \sum_{\substack{\mathcal{P}^{m\to j}\\|E(\mathcal{P})|=d(m,j)}}w(\mathcal{P})}
\end{align}
depends only on the weights of the shortest paths connecting the initially occupied states to $i$ and $j$.

\begin{figure*}[h]
    \centering
    \includegraphics[width=0.9\linewidth]{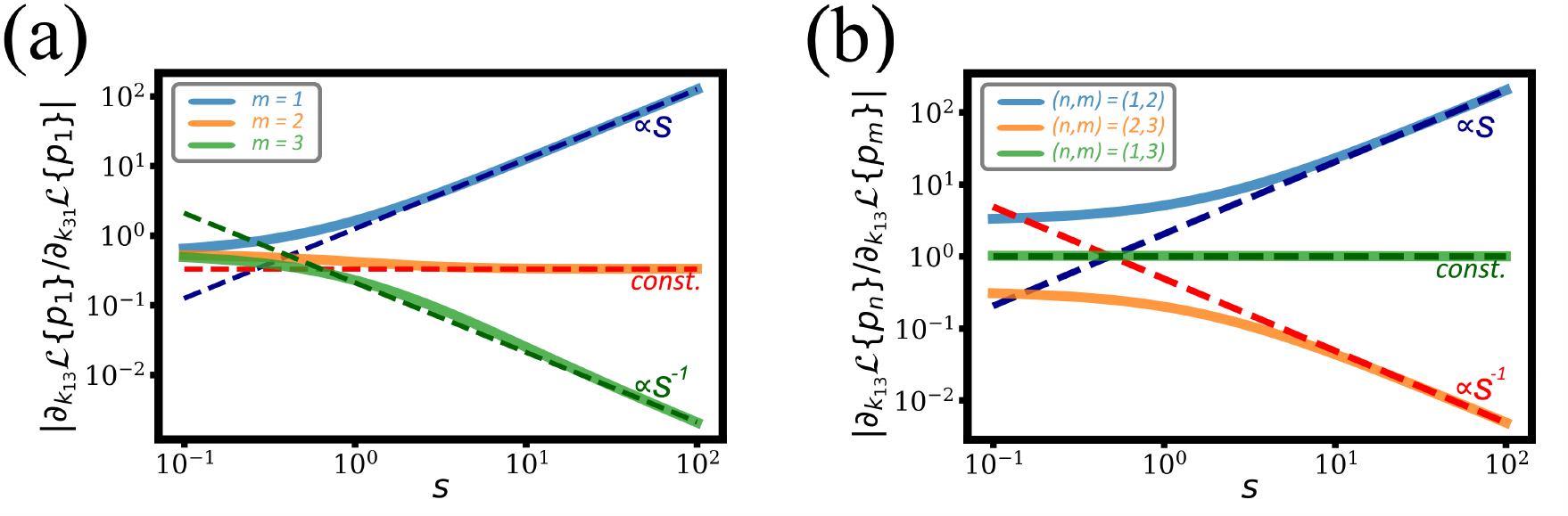}
    \caption{Illustration of the short time $s\to\infty$ scaling of the response ratios for the example network in Fig.~\ref{fig:mutual_linearity_illustration}. (a) The ratio $\frac{\partial_{k_{13}}\mathcal{L}\{p_1\}}{\partial_{k_{31}}\mathcal{L}\{p_1\}}(s)$ for different initial positions $m\in\{1,2,3\}$ (solid lines) admits three possible asymptotic scalings depending on the graph distance from $m$ to $1$ and $3$ (dashed lines). (b) The ratio $\frac{\partial_{k_{13}}\mathcal{L}\{p_n\}}{\partial_{k_{13}}\mathcal{L}\{p_m\}}(s)$ (solid lines) is determined by the shortest distances from $1$ and $3$ to $n$ and $m$ for the short time
    asymptotics (dashed lines).}
    \label{fig:mutual_linearity_asymptotics}
\end{figure*}

We note that $\Delta_{ij}\in\{-1,0,1\}$. This follows from the fact that if both transitions $i\to j$ and $j\to i$ are allowed, the distances of $i$ and $j$ from any vertex can differ by at most one. \new{The ratio of the occupation probability in Laplace space upon variation of the rates of two opposite transitions is shown in Fig.~\ref{fig:mutual_linearity_asymptotics}(a) for the example network from Fig.~\ref{fig:mutual_linearity_illustration} for different initial positions (solid lines) illustrating the corresponding asymptotic scalings (dashed lines).} 
Using the fact that the time integral of a function $g(t)$ corresponds to a multiplication with $s^{-1}$ in Laplace space,
\begin{align}
    \mathcal{L}\Big\{{\integ{0}{t}{t'}g(t')}\Big\}(s)=\frac{1}{s}\Laplace{g},
\end{align}
and the expression of the derivative in Laplace space, the three possible values of $\Delta_{ij}$ can be related to three universal short-time relations between the responses to the forward and backward rates:
\begin{align}
    \partial_{k_{ij}}p_n(t)\overset{t\to0}{\simeq}- C_{ij}
    \begin{cases}
        \displaystyle \int_0^t dt'\,\partial_{k_{ji}}p_n(t'), & \Delta_{ij}=-1,\\\qquad\partial_{k_{ji}}p_n(t), & \Delta_{ij}=0,\\\qquad\partial_{k_{ji}}\dot p_n(t), & \Delta_{ij}=1,
    \end{cases}
\end{align}
where in the last case, we used $\partial_{k_{ji}}p_n(0)=0$.

Similarly, for the susceptibility, the following is obtained:
\begin{align}
    \Laplace{\chi_{n,m}^{(i,j)}}\overset{s\to\infty}{\simeq}D_{n,m}^{(i,j)}\, s^{\Delta_{n,m}^{(i,j)}},\label{eq:short_time_chi_Laplace}
\end{align}
where
\begin{align} 
    \Delta_{n,m}^{(i,j)}&:=\min(d(i,m),d(j,m))-\min(d(i,n),d(j,n)). 
\end{align}
The prefactor in Eq.~(\ref{eq:short_time_chi_Laplace}) is of the shape $D_{n,m}^{(i,j)}:=\frac{d_n^{(ij)}}{d_m^{(ij)}}$, with the definition
\begin{align}
    d_n^{(ij)}=\begin{cases}
        \hspace{1.5cm}\sum_{\substack{\mathcal{P}^{j\to n}\\|\mathcal{P}|=d(j,n)}}w(\mathcal{P})&\text{if}~~{d(j,n)< d(i,n)}\\ \sum_{\substack{\mathcal{P}^{j\to n}\\|\mathcal{P}|=d(j,n)}}w(\mathcal{P})-\sum_{\substack{\mathcal{P}^{i\to n}\\|\mathcal{P}|=d(j,n)}}w(\mathcal{P})&\text{if}~~{d(j,n)= d(i,n)}\\ \hspace{1.3cm}-\sum_{\substack{\mathcal{P}^{i\to n}\\|\mathcal{P}|=d(i,n)}}w(\mathcal{P})&\text{if}~~{d(j,n)> d(i,n)}
    \end{cases}
\end{align}
In particular, $D_{n,m}^{(i,j)}$ depends only on the weights of the shortest paths connecting $i$ or $j$ to the corresponding states. \new{Fig.~\ref{fig:mutual_linearity_asymptotics}(b) 
shows for the example network in Fig.~\ref{fig:mutual_linearity_illustration}
that the asymptotic scalings of the ratios (dashed lines) are determined by the shortest path distance for different occupation probabilities in Laplace space (solid lines).}

Using the fact that the Laplace transform of higher order derivatives of a function $g(t)$ is given by
\begin{align}
    \Laplace{g^{(n)}}=s^n\Laplace{g}-\sum_{m=1}^ng^{(n-m)}(0)s^{m-1},
\end{align}
in time domain, we find the following short-time relations for $\Delta_{n,m}^{(i,j)}\geq0$:
\begin{align}
    \partial_{k_{ij}} p_n(t)\overset{t\to0}{\simeq} D_{n,m}^{(i,j)}~\partial_{k_{ij}} p_m^{\new{(\Delta_{n,m}^{(i,j)})}}(t),
\end{align}
since the lower-order derivatives are all zero for $t=0$ (Eq.~(\ref{eq:derivatives_vanish})). \new{Here, the superscript $(\Delta_{n,m}^{(i,j)})$ denotes the $\Delta_{n,m}^{(i,j)}$-th derivative of $p_m(t)$}. \new{The case $\Delta_{m,n}^{(i,j)}\geq 0$ follows directly by exchanging $n$ and $m$. We nevertheless state it explicitly for convenience}:
\begin{align}
    \partial_{k_{ij}} p_m(t)\overset{t\to0}{\simeq} D_{m,n}^{(i,j)}~\partial_{k_{ij}} p_n^{\new{(\Delta_{m,n}^{(i,j)})}}(t).
\end{align}

Thus, the short-time response of the network is governed entirely by the graph geometry: only the distances between states and the weights of the corresponding shortest paths enter the leading dynamics.

\section{Mutual linearity of hitting time densities}
\label{sec:fpt}

The hitting or first-passage time to state $n^\ast$, defined as $\tau:=\inf\{t>0~|~X_t=n^\ast\}$, is the random variable describing the time it takes for the process to reach state $n^\ast$. Here, we choose $n^*=N$, without loss of generality. Let $f_m(t)$ be the hitting time density to $N$ with the initial condition $p_n(0)=\delta_{nm}$ with $m\in\{1,...,N-1\}$. Since we are only interested in the first time that this state is reached, $N$ can be chosen as an absorbing state, i.e., we are considering the stochastic process on the graph without outgoing edges from $N$:
\begin{align}
    \mathcal{G}^{\setminus N}=(V(\mathcal{G}),~E(\mathcal{G})\setminus\{(N,i)~|~i\in\{1,...,N-1\}\}),
\end{align}
with the same edge-weights for the remaining edges. Then, if $X_{t_1}=N$, $X_{t_2}=N$ for all $t_2\geq t_1$, meaning that the process remains in this state once it is reached. Hence, the following holds:
\begin{align}
    F_m(t)&=\mathbb{P}(\tau\leq t~|~X_0=m)\\&=1-\sum_{i=1}^{N-1}p_i(t),
\end{align}
where $F_m(t)$ is the distribution function of $\tau$, corresponding to the probability that $\tau\leq t$, and the probability density $f_m(t)=\dot{F}_m(t)$ is given by
\begin{align}
    f_m(t)&=-\sum_{n=1}^{N-1}\dot{p}_n(t).
\end{align}
Employing Eq.~(\ref{eq:prob_density_Laplace_space}), the Laplace transform of $f_m(t)$ can also be expressed in combinatorial terms:
\begin{align}
    \Laplace{f_m}&=1-s\sum_{n=1}^{N-1}\Laplace{p_n}\\ &=1-\sum_{n=1}^{N-1}\frac{s\sum_{\mathcal{F}^{s, m\to n}_{[n,*]}}w(\mathcal{F})}{\sum_{\mathcal{T}^{s}_{[*]}}w(\mathcal{T})}\\ &=\frac{\sum_{\mathcal{T}^{s}_{[*]}}w(\mathcal{T})-s\sum_{n=1}^{N-1}\sum_{\mathcal{F}^{s, m\to n}_{[n,*]}}w(\mathcal{F})}{\sum_{\mathcal{T}^{s}_{[*]}}w(\mathcal{T})}\\ &=\frac{\sum_{n=1}^{N-1}k_{nN}\sum_{\mathcal{F}^{s, m\to n}_{[n,*]}}w(\mathcal{F})}{\sum_{\mathcal{T}^{s}_{[*]}}w(\mathcal{T})},\label{eq:fpt_density_Laplac_space}
\end{align} 
using the initial condition $p_m(0)=1$. 

Thus the Laplace transform of the hitting time densities can also be expressed in combinatorial terms, similar to the occupation probabilities (Eq.~(\ref{eq:prob_density_Laplace_space})). Therefore, analogous expressions \new{to Eqs.~(\ref{eq:ratio_k_ij_k_ji}) and (\ref{eq:def_chi})} for the relation of the hitting time densities under variation of one transition rate $k_{ij}$ can be derived. To that end, we consider
\begin{align}
    \partial_{k_{ij}}\Laplace{f_n}=&-s\sum_{l=1}^{N-1}\partial_{k_{ij}}\Laplace{p_l}\\&=s\frac{\sum_{\mathcal{F}^{s, i\to n}_{[n,*]}}w(\mathcal{F})-\sum_{\mathcal{F}^{s, j\to n}_{[n,*]}}w(\mathcal{F})}{\sum_{\mathcal{T}^{s}_{[*]}}w(\mathcal{T})}\Laplace{p_i}\\=&\Big(\big(1-s\frac{\sum_{\mathcal{F}^{s, j\to n}_{[n,*]}}w(\mathcal{F})}{\sum_{\mathcal{T}^{s}_{[*]}}w(\mathcal{T})}\big)-\big(1-s\frac{\sum_{\mathcal{F}^{s, i\to n}_{[n,*]}}w(\mathcal{F})}{\sum_{\mathcal{T}^{s}_{[*]}}w(\mathcal{T})}\big)\Big)\Laplace{p_i}\\=&\big(\Laplace{f_j}-\Laplace{f_i}\big)\frac{\sum_{\mathcal{F}^{s,n\to i}_{[i,\ast]}}w(\mathcal{F})}{\sum_{\mathcal{T}^s_{[\ast]}}w(\mathcal{T})},
\end{align}
using Eq.~(\ref{eq:Laplace_partial_derivative_p_n}) to express the derivatives and Eqs.~(\ref{eq:prob_density_Laplace_space}) and (\ref{eq:fpt_density_Laplac_space}) for the last equality.
Hence, for the hitting time densities in Laplace space, one finds
\begin{align}
    \frac{\partial_{k_{ij}}\Laplace{f_n}}{\partial_{k_{ji}}\Laplace{f_n}}&=-\Laplace{\xi_{(i,j)}^{n,n}}\label{eq:mutual_linearity_FPT_Laplace_1}\\\frac{\partial_{k_{ij}}\Laplace{f_n}}{\partial_{k_{ij}}\Laplace{f_m}}&=\Laplace{\xi_{(i,i)}^{n,m}},
\end{align}
where in analogy to the susceptibility for the probability distributions, we define
\begin{align}
    \Laplace{\xi_{(i,j)}^{n,m}}:=\frac{\sum_{\mathcal{F}^{s,n\to i}_{[i,\ast]}}w(\mathcal{F})}{\sum_{\mathcal{F}^{s,m\to j}_{[j,\ast]}}w(\mathcal{F})}.
\end{align}
The following properties again follow immediately from the definition:
\begin{align}
    \Laplace{\xi_{(i,j)}^{n,m}}=\Big(\Laplace{\xi_{(j,i)}^{m,n}}\Big)^{-1},~\Laplace{\xi_{(i,l)}^{n,k}}\Laplace{\xi_{(l,j)}^{k,m}}=\Laplace{\xi_{(i,j)}^{n,m}},
\end{align}
and, since both the numerator and the denominator of $\Laplace{\xi_{(i,i)}^{n,m}}$ are sums over forests with $i$ as a root, they cannot contain the edge $(i,j)$, implying that
\begin{align}
    \partial_{k_{ij}}\Laplace{\xi_{(i,i)}^{n,m}}=0.
\end{align}
Hence, a mutual linearity relation also holds for the Laplace transformed hitting time densities:
\begin{align}
    \Laplace{f_n}=\Laplace{g^{n,m}_{(i,j)}}+\Laplace{\xi_{(i,i)}^{n,m}}\Laplace{ f_m}\label{eq:mutual_linearity_FPT_Laplace}
\end{align}
and in the time domain:
\begin{align}
    f_n(t)=g^{n,m}_{(i,j)}(t)+(\xi_{(i,i)}^{n,m}\ast f_m)(t),
\end{align}
where $g^{n,m}_{(i,j)}$ and $\xi_{(i,i)}^{n,m}$ are both independent of $k_{ij}$.

It is then analogous to the derivation in Sec.~\ref{sec:asymptotics1} to obtain the short-time scalings for the susceptibility, from considering the dominant scaling for $s\to\infty$:
\begin{align}
    \Laplace{\xi_{(i,j)}^{n,m}}\overset{s\to\infty}{\simeq}C_{ij}^{nm}s^{d(m,j)-d(n,i)},
\end{align}
where the prefactor is defined as
\begin{align}
    C_{ij}^{nm}:=\frac{\sum_{\substack{\mathcal{P}^{n\to i}\\|\mathcal{P}|=d(n,i)}}w(\mathcal{P})}{\sum_{\substack{\mathcal{P}^{m\to j}\\|\mathcal{P}|=d(m,j)}}w(\mathcal{P})}.
\end{align}
In the time domain, this translates into the leading order behavior in $t$:
\begin{align}
    \partial_{k_{ij}}f_n(t)\overset{t\to 0}{\simeq}-C_{ij}^{nn}\begin{cases}
        \integ{0}{t}{t'}\partial_{k_{ji}}f_n(t')&\text{if}~~d(n,j)=d(n,i)-1\\~~~\partial_{k_{ji}}f_n(t)&\text{if}~~d(n,j)=d(n,i) \\~~~ \partial_{k_{ji}}\dot{f}_n(t)&\text{if}~~d(n,j)=d(n,i)+1.
    \end{cases} 
\end{align}
Again, only three cases are possible. Moreover, for \new{$d(n,i)\leq d(m,i)$} we have:
\begin{align}
    \partial_{k_{ij}}f_n(t)\overset{t\to 0}{\simeq}C_{ii}^{nm}\partial_{k_{ij}}f_m^{\new{(d(m,i)-d(n,i))}}(t), 
\end{align}
with the analogous relation obtained by exchanging $m$ and $n$:
\begin{align}
    \partial_{k_{ij}}f_m(t)\overset{t\to 0}{\simeq}C_{ii}^{mn}\partial_{k_{ij}}f_n^{\new{(d(n,i)-d(m,i))}}(t).
\end{align}

\section{Conclusion}

In this work, we have derived explicit expressions
for the Laplace transforms of the transient probability and hitting time distribution for a stochastic process on a Markov network with time-homogeneous rates as sums over the weights of spanning trees and spanning forests in an augmented graph\new{, compare Eq.~(\ref{eq:prob_density_Laplace_space}) and Eq.~(\ref{eq:fpt_density_Laplac_space}), respectively.} For the probability distribution, this can be seen as an extension of the Markov chain tree theorem (Eq.~(\ref{eq:MCTT})) to the non-stationary case. Our explicit expressions allowed us to give a graph theoretic proof for the mutual linearity relation under variation of a single transition rate for the Laplace-transformed occupation probabilities \new{(Eq.~(\ref{eq:mutual_linerarity_Laplace}))} that Zheng and Lu showed at the trajectory level~\cite{zheng2026mutual}. Our approach also resulted in an explicit formula for the susceptibility \new{(Eq.~(\ref{eq:def_chi}))}. For the long-time limit, we have shown that our expressions recover the results for the stationary distribution obtained by Bebon and Speck (\new{Eq.~(\ref{eq:chi_Bebon_Speck})}). 
For the short-time limit, we have shown that the leading order contribution is determined by minimal path distances from the initial distribution (\new{Eq. (\ref{eq:short_time_chi_Laplace})}).
\new{This is a well-known result in the theory of Markov networks, including the analogous result for the short-time scaling of hitting times \cite{li2013mechanisms,li2014pathway}.
In particular, these results have been verified experimentally~\cite{thorneywork2020direct}
and can be used as a method to detect hidden intermediate states in the network~\cite{van2022thermodynamic,maier2025observed}.}
Finally, based on their similar combinatorial structure in Laplace space, we have shown that mutual linearity also extends to hitting time densities \new{(Eqs.~(\ref{eq:mutual_linearity_FPT_Laplace_1}-\ref{eq:mutual_linearity_FPT_Laplace}))}.

While the Markov chain tree theorem has been known and applied for several decades~\cite{schnakenberg1976network,zia2007probability,gunawardena2012linear,nam2022linear}, the extension to hitting times is relatively recent~\cite{nam2025algebraic,voits2026emergence}, and a similar extension to non-stationary probabilities as in Eq.~(\ref{eq:prob_density_Laplace_space}) does not appear to have been considered explicitly in this combinatorial form. In the future, studying these combinatorial expressions in more detail may allow one to re-derive other known results and even prove new properties of Markov networks under relatively general conditions. 
Regarding applications, the graph theoretic structure and the mutual linearity following from it impose mathematical constraints on general Markov networks and are therefore of high relevance in optimized systems such as biochemical networks~\cite{tkavcik2025information}.
In principle, one can envision to use mutual linearity
as a design principle \new{or as a fundamental constraint} for networks with desired or even optimal information processing \cite{owen2020universal,avanzini2023circuit, floyd2025limits,floyd2025local}.

\new{Interestingly, the stationary reverse-edge response relation in Eq.~(\ref{eq:stationary_reverse_edge_response}) can also be derived from a framework for dynamical fluctuation-response relations established in \cite{aslyamov2026dynamical}, where it appears as a special instance of a more general theory also allowing for arbitrary integrated observables and time-dependent transition rates. Whether other results on mutual linearity can likewise be generalized to the non-autonomous case remains an open question for future research.  

Finally we note that the combinatorial treatment of the resolvent of Markov processes bears some formal resemblance to diagrammatic methods in quantum field theory, such as the canonical forest formula, which is used to reorganize coupled field-theoretic expressions into sums over marked trees \cite{gurau2009tree}. However, quantum field theory is fundamentally different in the interpretation of the underlying objects and combinatorial expansions, which leads to different types of algebraic structures and cancellations in the corresponding summations.
Moreover, the diagrammatic expansions
in Feynman graphs are determined less by system size and more by the order of
the perturbation theory. In the future, it may nevertheless be very fruitful to explore whether some of the combinatorial results
developed here for classical Markovian networks carry over to 
tree-based formulations of quantum field theory, or vice versa.}

%
%

\ack{JBV thanks the German Academic Scholarship Foundation (Studienstiftung des Deutschen Volkes) for support. We also acknowledge support by the Max Planck School Matter to Life funded by the Dieter Schwarz Foundation and the Max Planck Society.}

\roles{J B Voits \orcidlink{0009-0006-2650-6968}: Conceptualization, Investigation, Methodology, Writing – original draft, Writing – review \& editing (equal)

\noindent U S Schwarz \orcidlink{0000-0003-1483-640X}: Project administration, Supervision, Writing – review \& editing (equal)
}

\data{All data that support the findings of this study are included within the article.}





\providecommand{\noopsort}[1]{}\providecommand{\singleletter}[1]{#1}%
\providecommand{\newblock}{}

\end{document}